\documentclass[twocolumn,pra,showpacs]{revtex4}

\usepackage{amsmath,amssymb}
\usepackage{graphicx}

\begin{document}

\title{The short-range three-body phase and other issues impacting the observation of Efimov physics 
in ultracold
  quantum gases}  
\author{J. P. D'Incao}
\affiliation{Department of Physics and JILA, University of Colorado, Boulder, Colorado 80309}
\author{Chris H. Greene}
\affiliation{Department of Physics and JILA, University of Colorado, Boulder, Colorado 80309}
\author{B. D. Esry}
\affiliation{Department of Physics, Kansas State University,
  Manhattan, Kansas 66506}  

\begin{abstract}
We discuss several issues important for experimentally
observing Efimov physics in ultracold quantum gases.
By numerically solving the three-boson Schr\"odinger equation 
over a broad range of scattering lengths and energies, and by including
model potentials with multiple bound states, we address the complications of 
relating experimental observations to available analytic expressions.
These more realistic potentials introduce features that can mask
the predicted Efimov physics at small scattering lengths.  They also allow
us to verify that positive and negative scattering lengths are
universally connected only across a pole, not across a zero.
Additionally, we show that the spacing between Efimov features
for the relatively small scattering lengths accessible experimentally 
fail to precisely follow the geometric progression expected for Efimov physics. 
Finally, we emphasize the importance of the short-range three-body 
physics in determining the position of Efimov 
features and show that theoretically reproducing two-body physics 
is not generally sufficient to predict three-body properties quantitatively.
\end{abstract} 
\maketitle 

\section{Introduction}

The importance of three-body collisions in determining the lifetime and stability
of ultracold atomic gases has been underscored in recent experiments~\cite{ExpGeneral}. 
Near a Feshbach resonance,
for instance, the two-body $s$-wave scattering length $a$ can be made
much larger than the characteristic range $r_{0}$ of the interatomic
interactions and, due to the strong dependence of the
ultracold three-body collision rates on $a$ \cite{ScaLenDep}, the loss
processes can dominate, burning the sample completely or, in other
cases, leading to unexpected stability when such collisions are
suppressed \cite{LLM}. In addition to these practical considerations,
fundamental effects associated with Efimov physics also
appear in the strongly interacting limit  $|a|\gg r_{0}$.
In the early 1970's Vitaly Efimov predicted~\cite{Efimov}, in
the context of nuclear physics, the existence of a large number of
exotic weakly bound trimers that appear whenever $|a|\gg r_{0}$.
Only recently, though,
has the first experimental evidence of such Efimov states been
obtained in ultracold gases \cite{Rudi}, representing a
landmark step for a new generation of 
experiments designed to explore the rich world of few-body physics. 

In ultracold quantum gases, Efimov physics is revealed through its influence
on collisions rather than through observation of the states themselves.
In this case, Efimov physics can be observed by measuring the time
evolution of the atomic or molecular densities and extracting 
the collision rates~\cite{Rudi}. 
Efimov physics appears as a series of minima or maxima in these 
rates as a function of $a$ with the features separated by
the multiplicative factor $e^{\pi/s_{0}}$, where $s_{0}=1.00624$ for
identical bosons but can vary greatly for other
systems~\cite{EfimovOb}. Many of these features are 
associated with the creation of a new Efimov state in the system.  

In this paper we discuss some of the issues that must be considered
when attempting to establish connections between analytical
predictions for Efimov physics 
(see, for instance, the reviews \cite{BraatenReview,BraatenReview2} 
and Ref.~\cite{ScaLenDep})
and experimental data. These
analytic expressions are derived as expansions about $T$=0 
and  $|a|/r_0=\infty$, typically including only the leading term of the expansion.  
The limitations of these expressions have been explored both in 
temperature~\cite{Limits} and in scattering length~\cite{Limits,Manifestation}.  
In particular, it has been shown that the 
$|a|\gg r_0$ limit is approached only very slowly for
many key properties of three-body systems, leading to complications 
in applying the analytic expressions to experiment since,
unfortunately, experiments cannot easily access this limit.
Appropriately corrected analytical formulas, although potentially
universal themselves, are not yet known~\cite{BraatenReview,BraatenReview2}. 
Perhaps surprisingly, another class of complications arises from the fact
that some experiments --- even though they are performed at seemingly
``ultracold'' temperatures --- actually lie outside the range of
validity~\cite{Limits} of the zero temperature analytical expressions currently
available.  Finally, there are still other complications
that can arise for realistic interatomic potentials that have not been
incorporated into the analytical expressions.  These complications include, for instance, the finite 
range of the potentials, the existence of higher angular momentum two-body states, and the 
multichannel nature of the two-body interaction.

While progress is being made to find more general analytic 
expressions~\cite{JonsellEPL,BraatenPRA,Koehler,Stoof,Platter,HammerPRA,PetrovFeshbach},
direct numerical solution of the Schr\"odinger equation currently
remains the most reliable approach for non-zero collision energies, finite range potentials, 
and finite $a$.  We present our numerical results for 
three-boson recombination, $B$+$B$+$B$$\rightarrow$$ B_{2}$$+$$B$,
using the adiabatic hyperspherical representation~\cite{HeRecomb,Limits}
for a wide range of collision
energies.  Our calculations show that if the
$|a|\gg r_{0}$ condition for Efimov physics is not fully satisfied,
then the spacing between Efimov features can be quite far from
the predicted $e^{\pi/s_0}$. Moreover, we show that if the two-body
model has several bound states, then additional resonant effects can
appear, complicating the interpretation of Efimov features. We also 
demonstrate that if $a$ is changed from $+\infty$ to $-\infty$ by
crossing $a=0$, then the relation between Efimov states that occur for
$a>0$ and $a<0$ is not universal, unlike the case when $a$ changes
instead through $|a|=\infty$ across the resonance~\cite{BraatenReview2}. 
To illustrate these points, we present numerical results for
recombination using model interactions chosen to match
the recent Innsbruck Cs experiment~\cite{Rudi}.    

Finally, we discuss the importance of the short-range three-body 
physics in determining the precise position of the Efimov features.  
In our work, this physics is parameterized by a short-range phase 
(also in Ref.~\cite{MacekRecomb}), whereas effective field theory 
treatments have various parameterizations generally equivalent to 
a momentum cutoff~\cite{Efimov,BraatenReview}.
This three-body parameter cannot, however, be determined from 
knowledge of the near-threshold two-body physics.  Consequently, 
fitting the near-threshold two-body observables~\cite{Koehler,Stoof} 
alone is not sufficient to accurately predict the position of the Efimov 
features, no matter how complete.  To make this point clear, we show 
that including a non-pairwise-additive three-body interaction~\cite{AxilrodTeller,Launay}, 
known to be present in every triatomic system, changes the positions of the 
Efimov features, which is equivalent to changing the short-range 
three-body phase~\cite{JPBDAMOPPoster}.  So, even if the ``exact'' 
two-body potentials were used instead of just the near threshold fit, 
the positions of the Efimov features could not be quantitatively predicted.

\section{Theoretical Method}

Our method for solving the three-body Schr\"odinger equation has been
detailed elsewhere~\cite{HeRecomb,Limits}, but we include a brief description here for completeness
and clarity.

\subsection{Adiabatic hyperspherical representation}

We solve the three-body Schr\"odinger equation using the
adiabatic hyperspherical representation~\cite{Macek68,HeRecomb}. After
the usual separation of the center-of-mass motion, the
three-body problem can be described by the hyperradius $R$ and five
hyperangles, denoted collectively by $\Omega$. The hyperangular part of the
Schr\"odinger equation determines the relative motion of the three bodies, and the
hyperradial part determines the overall size of the system.
The five angular coordinates are chosen to be the Euler angles ($\alpha$,
$\beta$, and $\gamma$), specifying the orientation of the plane
defined by the three particles relative to the space-fixed frame, plus
two hyperangles ($\theta$ and $\varphi$) defined as a modified version
of Smith-Whitten coordinates~\cite{HeRecomb,Smith}.

The Schr\"odinger equation in hyperspherical coordinates can be
written in terms of the rescaled wave function $\psi=R^{5/2}\Psi$ as (in atomic units)
\begin{equation}
\left[-\frac{1}{2\mu}\frac{\partial^2}{\partial 
R^2}+H_{ad}(R,\Omega)\right]\psi(R,\Omega)=E\psi(R,\Omega),
\label{schr}
\end{equation}
where $\mu$ is the
three-body reduced mass ($\mu=m/\sqrt{3}$ for three identical bosons
of atomic mass $m$) and $E$ is the total energy. The adiabatic Hamiltonian $H_{ad}$ in 
Eq.~(\ref{schr}) is given by
\begin{eqnarray}
H_{ad}(R,\Omega)=\frac{\Lambda^{2}+15/4}{2\mu R^2}+V(R,\theta,\varphi).
\label{Had}
\end{eqnarray}
\noindent
In this expression, $\Lambda^{2}$ is the hyperangular kinetic energy; and $V$,
the potential energy.  The eigenfunctions $\Phi_\nu$ of $H_{ad}$,
\begin{equation}
H_{ad}(R,\Omega)\Phi_{\nu}(R;\Omega) =U_{\nu}(R)\Phi_{\nu}(R;\Omega),
\label{poteq}
\end{equation}
form a complete, orthonormal basis at each $R$. Tthe total wave
function $\psi(R,\Omega)$ can thus be written as
\begin{equation}
\psi(R,\Omega)=\sum_{\nu}F_{\nu}(R)\Phi_{\nu}(R;\Omega),
\label{chfun}
\end{equation} 
\noindent
where $\nu$ is a collective index that includes all the quantum numbers necessary to identify each 
channel.  If the expansion in Eq.~(\ref{chfun}) includes the complete, countably-infinite set 
of $\Phi_\nu$, $\psi(R,\Omega)$, then this representation of $\psi$ is in principle
exact.  In practice, of course, the sum is truncated to a finite number of terms, but it can be 
extended
systematically to obtain essentially any desired level of accuracy.
The eigenvalues $U_{\nu}(R)$ of Eq.~(\ref{poteq}) give the potential curves.

Substituting Eq.~(\ref{chfun}) into the full Schr\"odinger equation (\ref{schr}) and
projecting out $\Phi_{\nu}$ gives the hyperradial Schr\"odinger equations --- a system of coupled
ordinary differential equations:
\begin{widetext}
\begin{equation}
\left(-\frac{1}{2\mu}\frac{d^2}{dR^2}+U_{\nu}(R)\right)F_{\nu}(R)
-\frac{1}{2\mu}\sum_{\nu'}
\left(2P_{\nu\nu'}(R)\frac{d}{dR}+Q_{\nu\nu'}(R)\right)F_{\nu'}(R)
=EF_{\nu}(R).
\label{radeq}
\end{equation}
\end{widetext}
The nonadiabatic coupling terms $P_{\nu\nu'}(R)$ and $Q_{\nu\nu'}(R)$ 
are generated by the hyperradial
dependence of the channel functions and are responsible for inelastic
transitions. They are defined as
\begin{equation}
P_{\nu\nu'}(R) =
\left<\!\!\!\left<\!\Phi_{\nu}(R)\left| \frac{d}{dR}\right |\Phi_{\nu'}(R)\!\right>\!\!\!\right>
\label{Puv}
\end{equation}
\noindent
and
\begin{equation}
Q_{\nu\nu'}(R) =
\left<\!\!\!\left<\!\Phi_{\nu}(R)\left| \frac{d^2}{dR^2}\right 
|\Phi_{\nu'}(R)\!\right>\!\!\!\right>.
\label{Quv}
\end{equation}
\noindent
The double brackets indicate integration only over the angular coordinates $\Omega$
and traces over any spin degrees of freedom, with the hyperradius $R$ fixed.

The $S$-matrix elements --- and thus cross
sections and rates --- are found from the solutions of Eq.~(\ref{radeq}) \cite{Aymar}.
The main focus of this work,
the three-body recombination rate $K_{3}$, is obtained from the
$S$-matrix as follows~\cite{OurFirstPRL,HeRecomb}:
\begin{equation}
K_3=\sum_{J,\pi}\sum_{i,f}\frac{192 (2J+1)\pi^{2}}{\mu k^4}|S^{J\pi}_{f\leftarrow i}|^{2},
\label{K3Def}
\end{equation}
where $k=\sqrt{2\mu E}$ is the incident hyperradial wave number, $i$
labels the initial continuum channel, and 
$f$ labels the final two-body channels.
This expression also reflects the fact that we use the total
orbital angular representation since the total orbital angular momentum $J$
is conserved.  Total parity $\pi$ is also a good quantum number
labeling the $S$-matrix.

\subsection{Two-body potential model}

The potential $V$ used in Eq.~(\ref{Had}) for the present calculations is a
sum of atom-atom interactions, $V=v(r_{12})+v(r_{31})+v(r_{23})$,
which is most appropriate for spin-stretched atoms. 
We take advantage of universality to use a
two-body model potential for $v(r_{ij})$ that is computationally convenient.
That is, the Efimov effect
and other low-energy three-body properties do not depend on the
details of the interatomic interaction, only on the scattering length
and the characteristic length scale $r_0$ of the potential. 
The primary simplification this allows is to reduce the number of
two-body ro-vibrational bound states from hundreds or thousands, for realistic 
triplet alkali potentials, to something more manageable.  

The two-body
potential model adopted here is 
\begin{equation}
v(r)=D\,{\rm sech}^2\left( \frac{r}{r_0}\right)
\label{TwoBodyPot}
\end{equation}
where $D$ is the potential strength. With this potential, we can
easily produce two-body systems with different values of the scattering
length and different numbers of bound states by changing $D$~\cite{OurFirstPRL,Manifestation}. 

\begin{figure}[htbp] 
\includegraphics[width=3.3in,angle=0,clip=true]{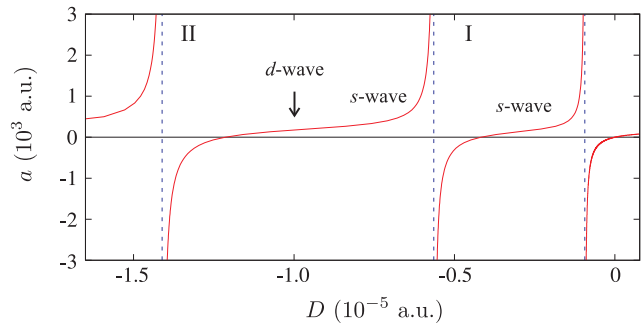}
\caption{Scattering length as a function of the potential depth for
  $v(r)$ from Eq.~(\ref{TwoBodyPot}).  There is
  one $s$-wave bound state in region I but there are two in region II.
  A $d$-wave bound state is also created in region II. 
 \label{Fig1}} 
\end{figure}

In Fig.~\ref{Fig1} we show the scattering length as a function of
$D$. The first pole in $a$ (as $D$ grows increasingly negative, 
starting from $D=0$) occurs when the first
$s$-wave bound state is formed.  The second pole in $a$ (as $|D|$
increases) occurs when a second $s$-wave bound state is formed.  
Our rate calculations cover regions
I and II. In addition to having one more $s$-wave
bound state than region I, region II also has a $d$-wave two-body state
that becomes bound at the value of $D$ indicated in the figure.  
As we will see below, this $d$-wave state
leads to resonance effects in the three-body recombination rate that
could be misinterpreted as an Efimov feature.  

\section{Results and Discussion}

The fact that three-body recombination can be the main loss
mechanism in ultracold atomic gases is what made Efimov physics observable
in the recent experiments of Kraemer {\it et al.}~\cite{Rudi}.
For $a>0$, the recombination rate at $T$=0 shows a series of minima,
resulting from interference between two different pathways~\cite{OurFirstPRL,MacekRecomb}. 
The rate can be conveniently written as
\begin{align}
K_{3}&= 67.1e^{-2\eta_{M}}
\!\!\left[
\sin^2[s_{0}\ln(\frac{a}{r_0})+\Phi_{M}]+\sinh^2\eta_{M}
\right]\frac{\hbar a^4}{m}\nonumber\\
&+16.8(1-e^{-4\eta_{M}})\frac{\hbar a^4}{m},
\label{K3apos}
\end{align}
including the contributions from both weakly- and deeply-bound 
molecules~\cite{OurFirstPRL,MacekRecomb,BraatenReview}.
For $a<0$, though, the mechanism that produces signatures of
Efimov physics is quite different and is
related to resonant transmission effects that occur when an
Efimov state is created~\cite{OurFirstPRL}. For $a<0$, the $T$=0 recombination rate 
is~\cite{BraatenReview}
\begin{eqnarray}
K_{3}=\frac{4590\sinh(2\eta_{P})}
{\sin^2[s_{0}\ln(|a|/r_{0})+\Phi_{P}]+\sinh^2{\eta_{P}}}\frac{\hbar
  |a|^4}{m}.
\label{K3aneg}
\end{eqnarray}
In these equations, $\Phi_{M}$ and $\Phi_{P}$ are short-range three-body 
phases that determine the positions of the minima and peaks,
respectively.  The parameters $\eta_{M}$ and $\eta_{P}$ were
introduced~\cite{BraatenReview} to characterize the probability of an
inelastic transition at small distances
to a deeply bound molecule. In practice, $\Phi$ and  
$\eta$ are used as fitting parameters since the short distance
behavior of realistic systems is not generally known.  We have
found~\cite{Limits} that Eqs.~(\ref{K3apos}) and (\ref{K3aneg}) are
good approximations at all collision energies $E$ and scattering lengths
$a$ in the threshold regime, i.e. $E\lesssim\hbar^2/2\mu_{\rm 2b}a^2$
where $\mu_{\rm 2b}$ is the two-body reduced mass. 

\subsection{Fitting experiment}
\label{ExptFit}

\begin{figure*}[htbp]
\includegraphics[width=6.in,angle=0,clip=true]{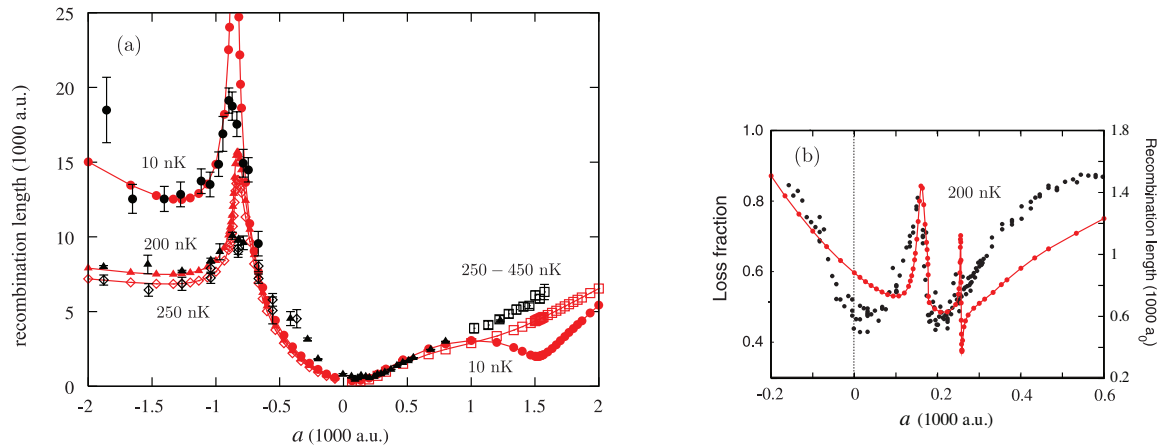}
\caption{(Color online) (a)
Comparison between $^{133}$Cs recombination length,  
$\varrho_{3}=(m K_{3}/\hbar\sqrt{3})^{1/4}$~\cite{OurFirstPRL},  obtained experimentally \cite{Rudi} 
(black symbols)
and our present thermally averaged results (red symbols). 
Filled circles and triangles were obtained at 10 and 200~nK, respectively;
open diamonds and squares were obtained at 250~nK and 250--450~nK,
experimentally, and at 250~nK in the present calculation. (b) 
Effect of the  $d$-wave resonance on the thermally averaged recombination length 
for small values of $a$ in region  II of Fig.~\ref{Fig1} in comparison to the observed loss
fraction from \cite{Rudi}.
}
\label{Fig2}
\end{figure*}

In Fig.~\ref{Fig2}(a) we show our numerical results for the thermally 
averaged~\cite{HeRecomb,Limits} recombination 
length $\varrho_3$ of $^{133}$Cs atoms, along with the experimental data obtained 
in the Innsbruck experiment 
\cite{Rudi}. Our data were generated for the same range of scattering lengths (in region I of 
Fig.~\ref{Fig1}) and temperatures
accessed experimentally, 
including the two leading-order contributions, $J^{\pi}=0^+$ and $2^+$, to Eq.~(\ref{K3Def}) 
\cite{Limits}.  
Even though we used a simple two-body model potential, 
our results for $a<0$ and $a>0$ reproduce the experimental data quite well.  
In our approach, the only adjustable parameter is $r_{0}$. We obtained $r_{0}$$\approx$100~a.u. by 
fitting 
the experimental peak position, $a\approx -850$~a.u., at $T$=10~nK.  We used this same $r_0$ to 
produce the curves for the other temperatures and for $a>0$.   We will discuss the validity of the 
latter in Sec.~\ref{Tuning}.

It is important to note that although our numerical value for $r_{0}$ 
happens to agree with the expected result for Cs
atoms --- namely, the van der Waal's length $r_{vDW}$~\cite{Julienne} --- it unfortunately does not 
imply that
we can predict the peak position for this or any other species.  The reason
is that our choice of model two-body potential Eq.~(\ref{TwoBodyPot}) determines the short-range
physics and has no particular relation to the correct short-range physics for three Cs atoms.  A 
different choice of model potential would lead to a different best fit value for $r_0$.  The model 
potential thus sets the values of $\Phi$ and $\eta$, preventing their use as fit parameters.  In 
fact, any finite range two-body model potential~\cite{Koehler,Stoof} will suffer from essentially 
the same problem, as we will discuss in more detail in Sec.~\ref{ShortRange}.  

The major difference between our results in Fig.~\ref{Fig2}(a) and
the experimental results lies at $a>0$.  As it turns out, we would not expect good agreement since 
our $a>0$ results use the same $r_0$ as the $a<0$ results.  As already noted, this point will be 
discussed in detail in Sec.~\ref{Tuning}, but the $a>0$ results in Fig.~\ref{Fig2}(a) do serve to 
illustrate the important role of temperature for attempts to observe the minimum.
Figure~\ref{Fig2}(a) shows, for instance, that at $10$~nK we do find a minimum at $a=1500$~a.u.. 
At 250~nK, however, strong contributions from the next partial wave, $J^{\pi}=2^{+}$~\cite{Limits}, 
mask the minimum.
Comparison with the experimental data shows that even if there were actually a minimum for Cs that 
matched our model, the experimental temperature was likely just too high to be able to see it.  For 
heavy atoms like Cs, the requirement of being in the threshold regime to see Efimov 
features~\cite{Limits} places a relatively severe limit on the maximum allowable temperature.

\subsection{Finite range potentials}

Analytic expressions for $K_3$ such as Eqs.~(\ref{K3apos}) and (\ref{K3aneg}) were derived as an 
expansion about $|a|\gg r_0$.  In practice, this was accomplished using zero-range two-body 
interactions that support at most one $s$-wave bound state.  Calculations performed using a finite 
range potential like Eq.~(\ref{TwoBodyPot}) produce deviations from these analytical predictions and 
introduces features not observable with these zero-range potentials.  One immediate consequence of a 
non-zero $r_0$ is that the minimum observed in the Cs experiment around $a=210$~a.u. is likely not 
related to Efimov physics since $a$ is not {\em much} larger than $r_0=r_{vDW}\approx$100~a.u.  This 
point was, in fact, already noted in Ref.~\cite{Rudi}.  We have found~\cite{Manifestation} that for 
most quantities the predictions of the zero-range model are recovered in a reasonably quantitative 
way only when $|a|$ is more than an order of magnitude larger than $r_0$.

In this range of scattering lengths, the recombination rate for finite range potentials can have 
structure that is completely independent of Efimov physics~\cite{JPBDAMOPPoster}.  This structure is 
generated by the presence of non-$s$-wave two-body states.  Figure~\ref{Fig2}(b)
shows an example of such structure due to a $d$-wave two-body state.
It shows our numerical results for $\varrho_3$, now
obtained for $a$ in region II (with two $s$-wave
bound states). These results display behavior similar to
the experimental observations, represented in this figure by the loss fraction reported
in Ref.~\cite{Rudi}. 
A minimum (or peak) in the recombination rate
does lead to a minimum (or peak) in the loss fraction, and
vice-versa, so the figures can be compared qualitatively. 
As the scattering length is decreased from its maximum in the figure, the two-body $d$-wave state 
can be visualized as approaching the threshold from above.  While above the threshold, the $d$-wave 
state is a two-body shape resonance; below threshold, it is, of course, a bound state.  Thus,
the first, sharp feature as $a$ is decreased, at
$a\approx 280$~a.u., is produced when the two-body shape resonance energy matches the three-body 
collision energy~\cite{NJP}.  Physically, this situation can be thought of as ro-vibrational 
relaxation of the two-body resonant state.  Two atoms collide, forming a $d$-wave two-body 
resonance, when the third atom collides with it, driving it down into the available $s$-wave bound 
states.  It is interesting to note that the total orbital angular momentum is $J=0$, whereby the 
relative angular momentum of the initial atom-dimer complex must also be a $d$-wave.  The final 
angular momentum of the bound dimer is zero as must be the relative atom-dimer angular momentum.  
There is thus considerable angular momentum and energy exchange at work near this feature.
The next broad feature at	
smaller $a$ occurs where this two-body resonance becomes a bound
state.  Being the highest-lying bound state, it receives the majority of the recombination.

It is worth noting that this $d$-wave $K_3$ resonance does not correspond to a resonance
in $a$ because it is an open channel two-body shape resonance --- not a resonance in the $s$-wave 
channel.  Higher
partial wave resonances in $a$ are typically due to higher order two-body
interactions coupling the incident $s$-wave to a closed channel resonances~\cite{Chin}.  The 
appearance of higher partial wave resonances at moderate scattering lengths is a generic feature 
that has been examined in detail for van der Waals interactions in Ref.~\cite{Gao}.  The two-body 
interaction for most any atoms will generally support not only $d$-wave bound states and resonances, 
but also much higher angular momentum states, potentially leading to considerably more structure 
near $a=0$.

We emphasize that the comparison in Fig.~\ref{Fig2}(b) is only an illustration of
the complications that can occur in realistic systems, especially in
the non-universal region of $|a|\approx r_0$.  A definitive
interpretation of the $a>0$ experimental data requires incorporating
considerably more information about the Cs-Cs interaction into our
model than we have.  Fortunately, much is known about the Cs-Cs interaction, and the scattering 
length has been calculated 
as a function of magnetic field with some confidence~\cite{Chin}. 
With that information, and the additional universal relations between $s$-wave and $d$-wave states
for alkali atoms \cite{Gao}, it should be possible to locate any higher partial wave open-channel
resonances to check whether features like the $d$-wave $K_3$ resonance
in Fig.~\ref{Fig2}(b) are possible.

Further complicating any effort to identify Efimov features
in this range of $a$, the global minimum in $K_3$ lies at $a\approx200$~a.u. --- even
without the $d$-wave resonance. 
The global minimum can be see in both Figs.~\ref{Fig2}(b) and (c), showing that recombination goes 
through a minimum when $a$ is tuned from $a\rightarrow\infty$ and $a\rightarrow-\infty$
through $a=0$. 

To further illustrate the effect of finite range two-body potentials, we show in
Fig.~\ref{Fig3} the recombination rate obtained for
fixed collision energies from $0.67\times10^{-4}$~nK 
up to $670$~nK and $|a|$ up to
$2\times10^6$~a.u. across regions I and II of Fig.~\ref{Fig1}.  These rates include only the $0^+$ 
contribution and are not thermally averaged, so the
indicated ``temperatures'' are the collision energies converted to Kelvin.
The rates in Figs.~\ref{Fig3}(a) and \ref{Fig3}(b) were obtained for
$a$ in region II and therefore have up to three contributions: 
recombination into the two possible $s$-wave states and into the
$d$-wave state.  The rates in Figs.~\ref{Fig3}(c) and \ref{Fig3}(d)
were obtained for $a$ in region I and represent recombination into the
single available $s$-wave bound state.  Figures~\ref{Fig3}(a) and (b) thus connect across a zero in 
the scattering length; (b) and (c), across a pole; and (c) and (d), across a zero.  Comparing the 
partial rates in Figs.~\ref{Fig3}(b) and (c)
for energies from 6.7$\times 10^{-2}$~nK up to 670~nK,
recombination into the lowest $s$-wave bound state is smooth across
the resonance where the second $s$-wave bound state is formed.
For lower energies this connection should also
be smooth, but will happen only for larger values of $a$.
Recombination into the second $s$-wave bound state in Fig.~\ref{Fig3}(b)
introduces a discontinuity in the total recombination rate at non-zero collision energies.  It 
would apprear from Fig.~\ref{Fig3}(b) that the appearance of the $d$-wave bound state near 
$a$=0 also introduces a discontinuity, but because the relative angular momentum between the
atom and dimer in the final state is $d$-wave, the Wigner threshold law guarantees that it will 
turn on smoothly.

Figure~\ref{Fig3}(a) shows that
recombination into the $d$-wave state actually dominates over the whole range of $a$ in this panel, 
which is not too surprising since it is the most weakly bound dimer state.
Moreover, the $d$-wave partial rate shows the expected Efimov features 
--- as do all of the partial rates in Fig.~\ref{Fig3}(a) --- 
emphasizing that the Efimov physics is determined
for $a<0$ by the initial three-body state and not the final
state~\cite{ScaLenDep,OurFirstPRL,MacekRecomb}.  This dependence is one way to understand the 
universality of $K_3$ for $a<0$ since the initial channel is universal while the final, deeply bound 
states are not.

One other issue evident from comparing the analytical curves in Fig.~\ref{Fig3} to the calculated 
ones is that the expected spacing between Efimov 
features does not hold between the two features nearest $a$=0.  In
Figs.~\ref{Fig3}(c) and \ref{Fig3}(d), the spacing between the first
two minima and first two peaks are 28.3 and 14.3, respectively, instead of the 22.7 expected for 
identical bosons. The spacing
between the next two minima and peaks are 23.4 and 21.6,
respectively.  Where the spacing can be defined in Figs.~\ref{Fig3}(a) and (b),
it follows the same pattern.
This sort of deviation for the spacing of the first few features
is important to be recognized by any experiment seeking to verify the 
logarithmic spacing of Efimov features.  At least in the near future, 
experiments will likely not be able to observe
more than a few of these features nearest to $a=0$, in view of the large range in $a$ and the low 
temperatures required~\cite{EfimovOb}. So, knowing the finite range corrections to the logarithmic 
spacing, and knowing whether those corrections are universal, are both critical.

\subsection{Tuning $a$ from $-\infty$ to $+\infty$}
\label{Tuning}

\begin{figure*}[htbp]
\includegraphics[width=7.in,angle=0,clip=true]{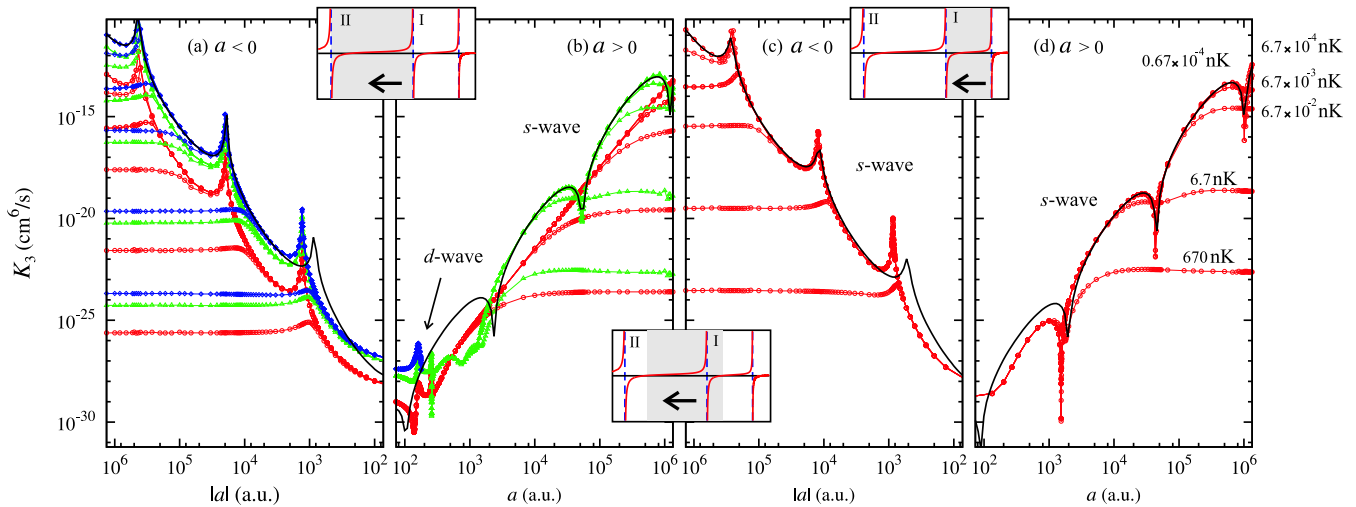}
\caption{(Color online) Recombination rate for $a$ covering regimes with
  different numbers of two-body bound states, as indicated above.  In (a) and (b)
  recombination has contributions from the deeply (red circles) and weakly (green triangles)
  bound $s$-wave states as well as from the $d$-wave state (blue diamonds).
  The lines without symbols are the fit of Eqs.~(\ref{K3apos}) and (\ref{K3aneg}).
  \label{Fig3}} 
\end{figure*}

In addition to the complications discussed above, 
fundamental issues were raised in Ref.~\cite{Rudi} since the
scattering length was tuned from $-\infty$ to $+\infty$ through $a=0$ 
and therefore through a non-universal region.   
In fact, $a$ was tuned through a series of narrow resonances.
While this was obviously the expedient choice experimentally, it
is not obvious that the $a<0$ and $a>0$ Efimov features should
retain their universal relation predicted from the zero-range model~\cite{BraatenReview},
\begin{equation}
\Delta\Phi=\Phi_M-\Phi_P=-1.53(3),
\label{PhaseDiff}
\end{equation}
or even that the short-range three-body phase on each side of a 
narrow resonance should be universally related, let alone the same.
Clearly, it is convenient if these relations hold since only one phase is
then required to predict the position of both minima
($a>0$) and peaks ($a<0$). So, the question
that naturally arises is whether there is a relation similar to
Eq.~(\ref{PhaseDiff}) between the peak and minimum positions when $a$
is tuned across a non-universal region ($a=0$). To explore this issue,
we analyze the relations between the numerical calculations shown in the panels of Fig.~\ref{Fig3}.
The fact that $K_3$ for $a<0$ and $a>0$ are universally related only across a {\it pole} in $a$, and 
the resulting consequences for the Cs experiment, have been raised in previous 
work~\cite{BraatenReview2,Koehler}.  
Here, however, we show via a counter-example that there is no general universal connection across a 
zero in $a$.  

Because of the slow approach of the system's properties to the $|a|\gg r_0$ predictions and the 
requirement of being in the threshold regime, the range of scattering lengths
and temperatures required to observe more than one feature in $^{133}$Cs, necessary
to determine $\Phi$ precisely from Eqs.~(\ref{K3apos}) and (\ref{K3aneg}), greatly exceeds current 
experimental capabilities. 
This fact underscores the advantages of using different atomic species to make the
required range of $a$ and $T$ much more reasonable \cite{EfimovOb}.

The three-body phases $\Phi_{P}$ and $\Phi_{M}$ were determined
from our numerical results in Fig.~\ref{Fig3} by independently fitting
Eqs.~(\ref{K3apos}) and (\ref{K3aneg}) to
the positions of the peaks and minima at our largest $|a|$ and lowest temperature in order to better 
satisfy the approximations under which those equations were derived.
The fit curves are indicated in Fig.~\ref{Fig3} by the
line without symbols.  From Figs.~\ref{Fig3}(b) and \ref{Fig3}(c),
which should be connected universally since they cross a pole in $a$, we 
obtain $\Delta\Phi_{bc}$=$-$1.553, in good agreement with 
Eq.~(\ref{PhaseDiff}). The other phase differences --- which
imply crossing a non-universal region --- 
are, however,  $\Delta\Phi_{ab}$=$-$1.053 and $\Delta\Phi_{cd}$=$-$1.381.
Thus, the Efimov physics for $a$$>$0 and $a$$<$0 do not appear
to be simply connected when separated by a non-universal region. 

We rationalize this non-universal behavior by
recognizing that small changes in the interactions should not substantially 
affect the short range physics, i.e. $\Phi$ and $\eta$.
Changing $a$ from positive to negative across a resonance
requires a much smaller change in the interactions than does crossing $a=0$,
so the former should lead to universal behavior.  The relative changes required in the depth of the 
potential $D$ when crossing each region
are clearly illustrated in Fig.~\ref{Fig1}.  Similarly, when controlling $a$ via a
Feshbach resonance, going from $a>0$ to $a<0$ 
requires smaller changes in the magnetic field when crossing a resonance than
when crossing $a=0$.  
It is not entirely clear, however, whether this argument generalizes directly
from our model potentials to realistic potentials with many bound states
since the relative change in such a realistic potential when crossing $a$=0
will also be small.  In principle, we can answer this question by simply continuing to increase $D$ 
and solving the problem numerically.  Unfortunately, we are not yet able to calculate the three-body 
recombination in a system with more than a handful of two-body bound states. Incidentally, we expect 
similar behavior for the other short-range three-body parameter $\eta$.  Specifically, we expect 
$\eta_P=\eta_M$ when $a$ is tuned across a resonance, but not when it is tuned across $a=0$.

In the Innsbruck experiment, $\Delta\Phi$ was found in the form of the ratio
\begin{equation}
\frac{a_{+}}{|a_{-}|}=\exp[{-(\Delta\Phi+\pi/2)/s_{0}}].
\end{equation}
Experimentally, $a_{+}$ located the first recombination maximum near $a=0$
for $a>0$; and $a_{-}$, the first peak for $a<0$.  The experimental
fit gave $a_{+}/|a_{-}|\approx 1.25(9)$~\cite{Rudi}.
Considering that the phases in $\Delta\Phi$ are defined as mod $\pi$, with $\Phi_{M}$ and $\Phi_{P}$ 
defined in the range $[0,\pi)$, the minimum and maximum values for $a_{+}/a_{-}$ are respectively 
0.00925 and 4.76385, this result agrees relatively well with the theoretical value of 0.96(3).
Theoretically, however, $\Delta\Phi$ is defined from the $|a|\gg r_0$ limit.
Our $\Delta\Phi_{bc}$, for instance, was determined by fitting in this limit 
and gives $a_+/|a_-|$=0.982.  The proper comparison with experiment, using
the first features in Figs.~\ref{Fig3}(b) and \ref{Fig3}(c), is difficult 
to make, though, since there is no clear maximum for small positive $a$ ---
save for the $d$-wave resonance --- to define $a_+$.  If there were,
it would likely be smaller than predicted by Eq.~(\ref{K3apos}) 
(compare with the solid line fit). On the other hand, $a_-$ would be
larger than predicted, giving $a_+/|a_-|$ smaller than the $|a|\gg r_0$ 
prediction --- assuming these feature shifts are universal.  We can also get
a sense of the quality of the agreement between theory and experiment by 
calculating $a_+/|a_-|$ from $\Delta\Phi$ across a non-universal region:
$\Delta\Phi_{cd}$, for instance, gives $a_+/|a_-|$=0.828; $\Delta\Phi_{ab}$ gives 0.598. 
Both of these values are nearly as close to the predicted value as the experiment, but
we know that the present ``agreement'' is accidental.
So, the fact that the experiment accessed the non-universal region together with
the fact that the minimum position ($a=210$~a.u.) is not firmly in the 
universal regime ($a\gg r_{0}$), make us believe that the
agreement between experiment and theory for $a_+/|a_-|$ is likely fortuitous.

\subsection{Three-body short-range physics}\label{ShortRange}

It should be clear by now that knowledge of the short-range three-body phase is essential to make 
quantitative theoretical predictions for ultracold three-body recombination of realistic systems --- 
or, indeed, for nearly any ultracold three-body process.  The three-body phase can be determined 
from essentially any ultracold three-body observable, and many examples have been discussed in 
Ref.~\cite{BraatenReview}.  It should be equally clear from the discussion of Fig.~\ref{Fig3} that 
fitting a two-body model potential to give the correct scattering length is not sufficient to 
determine the three-body phase.  Even a more complete fitting of the low-energy two-body 
physics~\cite{Koehler,Stoof} will not give the correct three-body phase.  In fact, quantitative 
predictions for real systems would not be possible even if the {\em complete} two-body interaction 
were included in the calculation.
The reason is simple: in real triatomic systems, there is a short-range, non-additive, purely three-
body interaction~\cite{AxilrodTeller,Launay}.
So, model calculations such as ours and those in Refs.~\cite{Koehler,Stoof},

that fit only two-body physics fundamentally cannot predict the positions of Efimov features 
quantitatively without the input of a known near-threshold three-body observable --- because of the 
three-body short-range physics.

To illustrate the effects of non-additive three-body interactions, we have performed calculations 
that include
one such correction. We have included the well-known three-body Axilrod-Teller potential 
\cite{AxilrodTeller} which, analogous to
the two-body van der Waals interaction, is the leading-order dispersion term describing the long-
range induced dipole-dipole-dipole interaction among
three neutral atoms. The Axilrod-Teller potential assumes the following form,
\begin{eqnarray}
W_{AT}(r_{12},r_{23},r_{13}) =
\gamma\frac{(1+3\cos\theta_{12}\cos\theta_{23}\cos\theta_{31})}{r_{12}^3r_{23}^3r_{31}^3},
\label{AxilrodTeller}
\end{eqnarray}
\noindent
where $r_{ij}$ are the interatomic distances and $\theta_{ij}=\cos^{-1}(\hat{r}_{ik}\cdot\hat{r}_{jk})$
are the inner angles of the triangle formed by the three atoms. In the above equation $\gamma$ is a 
positive quantity whose exact value depends on the details of the three-body system. Note that this
interaction can be attractive or repulsive depending upon the shape of the three atom system. It is 
also worth emphasizing
that the Axilrod-Teller interaction is just one of several short-range three-body interactions that 
can affect three-body
observables. In fact, much stronger corrections are expected due to purely three-atom exchange 
effects \cite{Launay}.
\begin{figure}[htbp]
\includegraphics[width=3.4in,angle=0,clip=true]{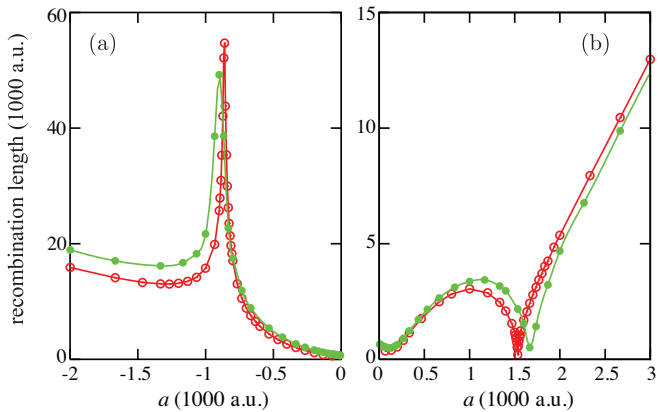}
\caption{(color online) Comparison of the (a) peak and (b) minimum position in the recombination 
length with the Axilrod-Teller interaction
 (green filled circles) and without it (red open circles).  These calculations demonstrate that the 
main effect of the inclusion
of the Axilrod-Teller potential is to move the position of the minimum and peak as shown above.}
\label{Fig4} 
\end{figure}

While $\gamma$ in Eq.~(\ref{AxilrodTeller}) normally reflects the details of the atoms, for the 
purposes of our model, we have chosen it to have the form
\begin{equation}
\gamma=\gamma_0 |D| r_0^9[\tanh(\frac{r_{12}}{r_0})\tanh(\frac{r_{23}}{r_0})\tanh(\frac{r_{31}}{r_0})]^3.
\end{equation}
This choice ensures that $W_{AT}$ scales as we tune the two-body scattering length, that it scales 
properly with $r_0$, and that the $r_{ij}^{-3}$ singularities are cut off near $r_{ij}=0$.  We have 
arbitrarily chosen a value for $\gamma_0$ that causes a variation of about 15\% in the adiabatic 
hyperspherical potentials
near the minimum.  This magnitude change is probably an underestimate, given that it has been shown 
that non-additive three-body terms make the minimum in the three-atom potential surface from 60\% to 
a factor of four deeper for alkali atom systems, compared to the pairwise additive potential 
surface~\cite{Hutson,HutsonLaunay,HutsonPRA}.

Figure~\ref{Fig4} shows the recombination length with and without the Axilrod-Teller term.
Our results in the figure demonstrate that
the minimum and peak positions are clearly shifted by the inclusion of the Axilrod-Teller term.
For $a<0$, besides changing the peak position, the Axilrod-Teller interaction also changes the width 
of the resonance peak
and the overall amplitude of recombination length. 
This result indicates that the coupling to deeply bound decay channels is also 
affected by the Axilrod-Teller interaction.  The analytic expressions in Eqs.~(\ref{K3apos}) and 
(\ref{K3aneg}) account for these effects through changes to the short-range parameters $\Phi$ and 
$\eta$.  

Our results in Fig.~\ref{Fig4}, therefore, demonstrate our initial statement. The inclusion of 
purely three-body
interactions such as the Axilrod-Teller --- or the three atom exchange --- are of crucial importance 
for accurately predicting
features related to Efimov physics. As a consequence, even though previous works have rather 
carefully fit
the two-body physics~\cite{Koehler,Stoof}, it is unlikely that they have predictive power in 
locating Efimov features.

\section{Summary}

Using essentially exact numerical solutions for a model two-body potential,
we have demonstrated the need for caution when applying existing analytic
expressions for three-boson recombination to experiment.  While they do
give considerable insight, their predictions for the first features near
$a$=0 are 
not quantitative and may even fail qualitatively due to the
complications arising from realistic potentials --- in particular, from their finite range.  
Moreover, care must be
taken to ensure that the temperature is truly in the threshold regime.
We have also
verified quantitatively that $a>0$ features are universally connected to $a<0$ features
only when $a$ is tuned through $|a|=\infty$ of a given resonance.

Finally, we have emphasized the importance of the short-range three-body phase in making 
quantitative predictions for real systems.  This point does not appear to be fully appreciated 
theoretically, but our example of the non-additive three-body term underscores the limitations of 
purely two-body models for actual triatomic systems and thus the need for a three-body parameter.  
Unfortunately, purely {\it ab initio} calculation of this three-body parameter seems, for now, out 
of reach for the systems of experimental interest.

\acknowledgments
We gratefully acknowledge a critical reading of an early version of this manuscript by R. Grimm's
group and their sharing their data.
This work was supported in part by the National Science Foundation and in part by the Air Force 
Office of Scientific Research.

\end{document}